\documentclass[twocolumn,showpacs,prb,citeautoscript]{revtex4}
\usepackage{graphicx}
\usepackage{dcolumn}
\usepackage{bm}
\usepackage{amsmath}
\usepackage{amssymb}
\usepackage{epsfig}

\begin{document}

\title{Crossed Andreev Reflection in Diffusive Contacts}
\author{A.\ Brinkman}
\author{A.A.\ Golubov}
\affiliation{Faculty of Science and Technology and MESA+ Institute for Nanotechnology,\\
University of Twente, 7500 AE, Enschede, The Netherlands}
\date{\today }

\begin{abstract}
Crossed Andreev reflection in multiterminal structures in the
diffusive regime is addressed within the quasiclassical
Keldysh-Usadel formalism. The elastic cotunneling and crossed
Andreev reflection of quasiparticles give nonlocal currents and
voltages (depending on the actual biasing of the devices) by
virtue of the induced proximity effect in the normal metal
electrodes. The magnitude of the nonlocal processes is found to
scale with the square of the barrier transparency and to decay
exponentially with interface spacing. Nonlocal cotunneling and
crossed Andreev conductances are found to contribute equally to
the nonlocal current, which is of relevance to the use of normal
metal-superconducting heterostructures as sources of entanglement.
\end{abstract}


\maketitle

\section{Introduction}
In standard Andreev reflection at a single normal
metal-superconductor interface an electronlike quasiparticle in
the normal metal can be transformed into a holelike quasiparticle
of opposite momentum \cite{Andreev}. When two normal metal ($N$)
electrodes or ferromagnets ($F$) are attached to a superconductor
($S$) at a distance from each other of the order of the coherence
length, two additional nonlocal processes are possible. During
elastic co-tunneling (EC), an electron is transferred from one
electrode to the other, while for crossed Andreev reflection (CAR)
an electron in one of the electrodes is transformed into a hole in
the other electrode \cite{Flatte,Deutscher}.

Bell-inequality experiments, quantum computation, and
teleportation of quantum states require quantum objects that are
entangled. Cooper pairs in superconductors are spin singlets and
are, therefore, suitable sources of entanglement. Crossed Andreev
reflection is a promising possibility for the creation of locally
separated entangled electrons
\cite{Recher,Chtchelkatchev,Chtchelkatchev2,Bayandin,Kawabata}.

Nonlocal transport properties have been seen experimentally in $NS$
\cite{Klapwijk,Chandra} and $FS$ heterostructures
\cite{Beckmann,Koren}. The microscopic origin of the effects of EC
and CAR, as well as possible ways in which these can be used for the
production of locally separated entangled quasiparticles, remain to
be understood theoretically.

Recently, theoretical studies of CAR have been performed for
various types of junctions \cite{Falci,Yamashita,Bignon,Feinberg,
Melin,Prada,Morten}. Modeling of nonlocal effects by means of
perturbation theory using an effective tunnel Hamiltonian
\cite{Falci,Bignon} has indeed provided EC and CAR, the signal
being, however, vanishingly small (of the order of the fourth
power in interface transparency). Consequent pioneering efforts to
include disorder \cite{Feinberg,Morten} and weak localization
\cite{Melin} have found enhanced effects, but still not to the
level of experimental observations, so that the question as to the
microscopic origin of CAR still remains open.

\begin{figure}[tbp]
\includegraphics [scale=0.42]{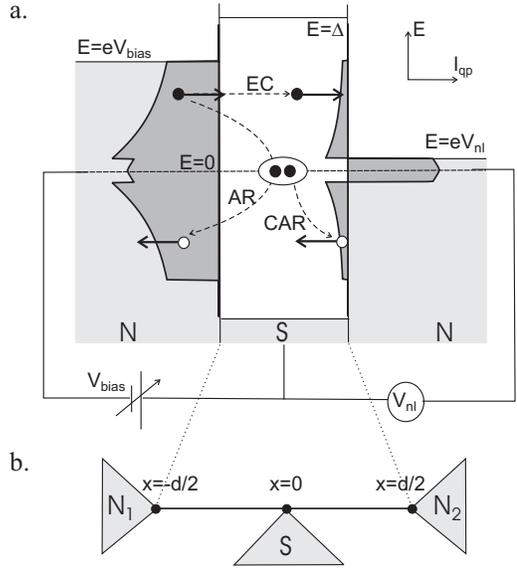}
\caption{(a) Schematic representation of the double-barrier $NSN$
structure, which is voltage biased accross the first interface. An
incoming electron from the left normal metal electrode, can
undergo three different processes that contribute to the current:
Andreev reflection (AR), elastic cotunneling (EC) and crossed
Andreev reflection (CAR). The depicted spectral currents that
result from these processes are the main result of this paper. (b)
Results have been obtained from a one-dimensional Keldysh-Usadel
quasiclassical Green's function calculation of the depicted
structure.} \label{schematic}
\end{figure}

In this paper, we provide insight into the underlying microscopic
mechanism of CAR by studying nonlocal transport by means of
quasiclassical kinetic theory. We present a mechanism in which CAR
exists by virtue of the proximity induced superconducting
correlations in the electrodes, for arbitrary barrier
transparency. The proximity effect is the essential new ingredient
in our model, giving a large contribution to CAR, of second order
in transparency, in contrast to the nonlocal effects of tunnel
Hamiltonian models. We show how CAR relates to the competing
process of co-tunneling. For ballistic transport, Andreev
reflection is understood most straightforwardly, but we have
modelled the additionally challenging case of diffusive transport,
as experimentally often is the case. Our model is of relevance to
the future design of experiments that are based on Andreev
entanglers.

\section{Quasiclassical model}
The most generic model system to study nonlocal effects in
superconducting structures is a three-terminal configuration
consisting of a quasi-1D superconducting wire of length $d$,
attached to normal reservoirs $N_{1,2}$ and a superconducting
reservoir $S$ which can be independently biased, see
Fig.~\ref{schematic}. This model is an extension to the earlier
approach by Volkov \textit{et al.} \cite{Volkov} that was used to
calculate nonequilibrium transport properties of two-terminal $N'NS$
contacts. The electrodes $N_{1,2}$ are weakly coupled to the wire,
while the reservoir $S$ is in good electrical contact with the wire
so that their electric potentials are equal. Morten \textit{et al.}
\cite{Morten} addressed in their circuit model the role of the
latter coupling strength between superconductor $S$ and wire. They
found nonvanishing nonlocal effects, but only in the case of weak
coupling. In the case of zero resistance between wire and $S$
(strong coupling), the circuit results coincide with the tunnel
Hamiltonian results of a vanishing nonlocal signal. However, in the
experiments often no barrier is present between wire and
superconductor. Hence, here we study the regime of good electrical
contact and equal potentials between $S$ and the wire.

We assume that transport is diffusive (scattering length being
smaller than other length scales) so that the quasiclassical
kinetic theory in the dirty limit can be applied. The
Keldysh-Usadel diffusion equation for the Green's function, in
Keldysh-Nambu space, in the absence of time-dependencies, magnetic
field and inelastic self-energy terms, can be written as
\begin{equation}
-\hbar \mathcal D \nabla \left(
{\mathord{\buildrel{\lower3pt\hbox{$\scriptscriptstyle \smile$}}
\over G}\nabla \mathord{\buildrel{\lower3pt\hbox{$
\scriptscriptstyle\smile$}} \over G}}\right) =\left[
iE{\mathord{\buildrel{ \lower3pt\hbox{$\scriptscriptstyle\smile$}}
\over \tau }_{3}+
\mathord{\buildrel{\lower3pt\hbox{$\scriptscriptstyle\smile$}}
\over \Delta},
\mathord{\buildrel{\lower3pt\hbox{$\scriptscriptstyle\smile$}}
\over G}} \right] , \label{KU}
\end{equation}
where
$\mathord{\buildrel{\lower3pt\hbox{$\scriptscriptstyle\smile$}}
\over \tau } _{3}=\left( {\begin{array}{*{20}c} {\hat \tau _3 } &
0 \\ 0 & {\hat \tau _3 } \\ \end{array}}\right)$, $\hat \tau
_3=\left( {\begin{array}{*{20}c} 1 & 0 \\ 0 & {-1}
\end{array}}\right)$,
\begin{equation*}
\mathord{\buildrel{\lower3pt\hbox{$\scriptscriptstyle\smile$}}
\over G} =\left( {\begin{array}{*{20}c} {\hat G^R } & {\hat G^K }
\\ 0 & {\hat G^A }
\end{array}}\right),\mathord{\buildrel{\lower3pt\hbox{$ \scriptscriptstyle\smile$}}
\over \Delta }=\left(
{\begin{array}{*{20}c} {\hat \Delta } & 0 \\ 0 & {\hat \Delta } \\
\end{array}}\right), \hat{\Delta}
=\left( {\begin{array}{*{20}c} 0 & \Delta \\ {\Delta ^* } & 0 \\
\end{array}} \right),
\end{equation*}
$E$ is the energy measured from the chemical potential, and
$\mathcal D$ is the diffusion constant. The current is given by
\begin{equation}
I = \frac{1}{{2eR_N }}\int {dE{\rm{Tr}}} \left[ {\hat \tau _3
\left( {\hat G^R \nabla \hat G^K  + \hat G^K \nabla \hat G^A }
\right)} \right], \label{current}
\end{equation}
where $R_N$ is the normal state resistance.

The quasiclassical modeling of the diffusive transport through
such a $NSN$ structure can be splitted in solving the retarded and
Keldysh parts of Eq. (\ref{KU}) respectively.

\subsection{The proximity effect}
In order to calculate the retarded part of the Green's function,
$\hat{G}^{R}$, it is convenient to use the standard $\theta
$-parametrization, ${ \hat{G}^{R}}(x)=\hat{\tau}_{3}\cos \theta
(x)+\hat{\tau}_{2}\sin \theta (x).$ The function $\theta (x)$ is a
measure of the superconducting correlations at a given point
within the structure and $\theta$ satisfies the Usadel equation
\cite{Usadel}
\begin{equation}
\mathcal D\frac{\partial ^{2}}{\partial x^{2}}\theta (x)+2iE\sin
\theta (x)=0. \label{Usa1}
\end{equation}

At the $NS$ interfaces at $x=\pm d/2$, the function $\theta (x)$ satisfies
the following boundary conditions \cite{KL}
\begin{eqnarray}
\gamma _{B}\xi _{N}\frac{\partial \theta _{N}}{\partial x} &=&\pm \sin
(\theta _{S}-\theta _{N}),\text{ \ \ }x=\pm d/2,  \label{KL1} \\
\gamma \xi _{N}\frac{\partial \theta _{N}}{\partial x} &=&\xi
_{S}\frac{
\partial \theta _{S}}{\partial x},\text{ \ \ }x=\pm d/2.  \label{KL2}
\end{eqnarray}%
where $\xi _{N,S}=\sqrt{D_{N,S}/2\pi T_{c}}$ are the coherence
lengths and $ D_{N,S}$ are the diffusion coefficients in $N$ and
$S$ respectively. The proximity effect parameters $\gamma$ and $
\gamma _{B}$ are defined as $\gamma _{B}=R_{B}/\rho _{N}\xi _{N}$
and $\gamma =\rho _{S}\xi _{S}/\rho _{N}\xi _{N}$, where $R_{B}$
is the interface resistance and $\rho _{N,S}$ are the
resistivities of the $N$ and $S$ metals. These parameters have a
simple physical meaning \cite{KL}: $\gamma$ is a measure of the
strength of the proximity effect between the $S$ and $N$ metals,
whereas $\gamma _{B}$ describes the effect of the interface
transparency. From here on, $\gamma _{B} \gg 1$ is assumed,
corresponding to a small barrier transparency.

Solutions to the proximity effect problem in diffusive junctions
have been extensively discussed in various regimes
\cite{Belzig,RevModPhys}. Generally, a minigap exists in $N$, of
the order of the Thouless energy. In the considered case of bulk
$N$ ($d_N \gg \xi_N$), the minigap vanishes. The quasiparticle
density of states is given by
$\operatorname{Re}G=\operatorname{Re}(\cos \theta )$ and the pair
density of states is defined as
$\operatorname{Re}F=\operatorname{Re} (\sin \theta )$. It is
straightforward to solve the Usadel equation (\ref {Usa1}) with
the boundary conditions (\ref{KL1}) and (\ref{KL2}) numerically.
The calculated densities of states $\operatorname{Re}G_{N}$ and
$\operatorname{Re} F_{N}$ at the $N$ side of an $NS$ interface are
shown in Fig. 2. It is seen that the quasiparticle spectrum in $N$
is gapless while strong superconducting correlations exist at low
$E$, described by $\operatorname{Re}F_{N}$. As will be shown
below, the existence of a nonzero $\operatorname{Re}F_{N}(E)$ at
the $ NS$ interface is an essential ingredient to our solution of
the nonlocal conductance in a diffusive $NSN$ structure.

\begin{figure}[tbp]
\includegraphics [scale=0.70]{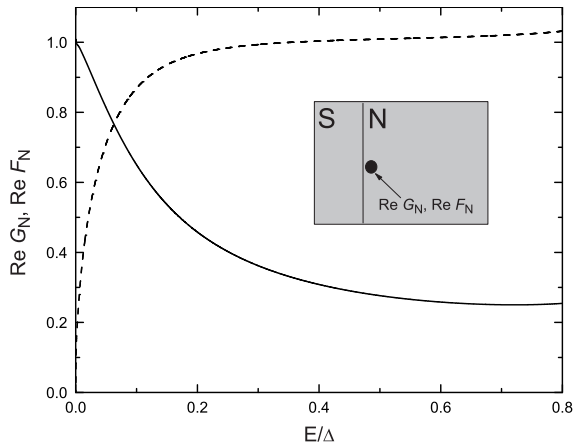}
\caption{Calculated density of states (Re$G_N$, dashed line) and
pair amplitude (Re$F_N$, solid line) as induced by a
superconductor into a normal metal, with $\protect\gamma_B$=5,
$\protect\gamma$=0.1, and $d_N \gg \protect\xi_N$. The functions
are plotted for the $N$-side of the interface, as indicated.}
\label{prox}
\end{figure}

\subsection{Distribution functions}
The Keldysh part of Eq.~(\ref{KU}) provides the distribution of
quasiparticles over energy. $\hat{G}^{K}$ can be parametrized as
$\hat{G}^{K}=\hat{G}^{R}\hat{f}-\hat{f}\hat{G}^{A}$, where the
distribution function $\hat{f}$ can be split into parts that are,
respectively, odd and even in energy,
$\hat{f}=f_{L}\hat{1}+f_{T}\hat{\tau}_{3}$. The kinetic equations
for the longitudinal ($f_{L}$) and transverse ($f_{T}$)
distribution functions have the form \cite{Belzig}
\begin{eqnarray}
\nabla \left( D_{T}\nabla f_{T}\right)
+\operatorname{Im}I_{S}\nabla f_{L}
&=&2f_{T}\Delta ,  \label{fT} \\
\nabla \left( D_{L}\nabla f_{L}\right)
+\operatorname{Im}I_{S}\nabla f_{T} &=&0 \label{fL},
\end{eqnarray}%
where $D_{T}=\left( \operatorname{Re}G\right) ^{2}+\left(
\operatorname{Re}F\right) ^{2}$, $ D_{L}=\left(
\operatorname{Re}G\right) ^{2}-\left( \operatorname{Im}F\right)
^{2}$ and $ \Delta $ is the gap inside the superconductor. $f_{T}$
and $f_{L}$ determine the quasiparticle and energy flow as can be
derived \cite{Belzig} from Eq. (\ref {current}), giving
respectively
\begin{eqnarray}
I_{qp} &=&\frac{1}{{2eR_{N}}}\int {dED_{T}\left( E\right) \nabla f_{T}\left(
E\right) }, \\
I_{L} &=&\frac{1}{{2eR_{N}}}\int {dED_{L}\left( E\right) \nabla
f_{L}\left( E\right) } \label{IL}.
\end{eqnarray}

The spectral supercurrent is given by
$\operatorname{Im}I_{S}=\frac{1}{8}{
\operatorname{Tr}}[{\hat{\tau}_{3}({\hat{G}^{R}\nabla
\hat{G}^{R}-\hat{G}^{A}\nabla
\hat{G}^{A}})}]=\operatorname{Im}F^{R}\operatorname{Re}F^{R}{\nabla
\chi }$, where ${ \chi }$ is the superconducting phase. We
consider the regime when superconductivity in the $S$-wire is not
influenced by the normal contacts $ N_{1,2}$, which is realized
when $\gamma \ll 1$ (large normal-state resistivity of $N$ metal
compared to that of $S$) \cite{RevModPhys}. Then, the product
$\operatorname{Im} F_{S}^{R}\operatorname{Re}F_{S}^{R}$ in $S$ is
nonzero only at a narrow energy range near $\Delta $. Since we are
interested only in the energy range $E<\Delta $ , this allows us
to neglect the terms with $\operatorname{Im}I_{S}\propto
\operatorname{Im} F_{S}^{R}\operatorname{Re}F_{S}^{R}$ in the
kinetic equations (\ref{fT}) and (\ref {fL}) at these energies,
and the equations for $f_{T}$ and $f_{L}$ decouple. The equation
for $f_{T}$ in the wire, consequently, becomes
\begin{equation}
D_{T}\frac{\partial ^{2}f_{T}}{\partial x^{2}}=2f_{T}\Delta
,\label{finaleq}
\end{equation}%
with the boundary conditions at the $NS$ interfaces given by
\cite{Brinkman}
\begin{equation}
D_{T}\gamma _{B}\frac{\partial }{\partial x}f_{T}=\pm
\operatorname{Re}F\operatorname{Re} F_{N}\left(
f_{T}-f_{TNi}\right)\label{finalbc}
\end{equation}
at $x=\pm d/2$, where $f_{TN}(\pm d/2)$ are the transverse
distribution functions in the normal reservours.

Note, that if the assumption $\gamma \ll 1$ is violated, the
density of states $\operatorname{Re}G$ in the wire becomes finite
at subgap energies. Then, the term
$\operatorname{Re}F\operatorname{Re} F_{N}$ in the above boundary
condition will be substituted by
$\operatorname{Re}F\operatorname{Re}
F_{N}+\operatorname{Re}G\operatorname{Re} G_{N}$. However, this
will not lead to any qualitative changes in our results and the
physical mechanism for the crossed Andreev transport remains the
same in this case.

Because of the small barrier transparency, the potential mainly
drops at the interfaces and we can assume the electric potential of
the\ $S$-reservoir and the wire to be zero, and the normal
electrodes $N_{1,2}$ to be in equilibrium with the potentials
$V_{1,2}$ respectively. Then, the distribution functions in
$N_{1,2}$ are
\begin{equation*}
f_{Ti,Li}=\frac{1}{2}\operatorname{tanh}\left(
\frac{E+eV_{i}}{2k_{B}T}\right) \mp
\frac{1}{2}\operatorname{tanh}\left(
\frac{E-eV_{i}}{2k_{B}T}\right) .
\end{equation*}%
The kinetic equation has the solution $f_{T}=Ae^{x/\xi }+Be^{-x/\xi }$%
\thinspace \thinspace\ where the coherence length $\xi $ is given by $\xi =%
\sqrt{D_{T}/2\Delta }$, which describes the conversion of
quasiparticle
current into supercurrent. This supercurrent is extracted by the $S$%
-reservoir at $x=0$.

\section{Results and discussion}
Solving Eq. (\ref{finaleq}) with the boundary conditions, Eq.
(\ref{finalbc}), the solution of $f_{T}$ in the case of two
symmetric $NS$ interfaces is given by
\begin{eqnarray}
A &=&N\frac{f_{T1}e^{-d/2\xi }\left( \gamma _{B}-N\right) +f_{T2}e^{d/2\xi
}\left( \gamma _{B}+N\right) }{e^{-d/\xi }\left( \gamma _{B}-N\right)
^{2}-e^{d/\xi }\left( \gamma _{B}+N\right) ^{2}},  \notag \\
B &=&N\frac{f_{T1}e^{d/2\xi }\left( \gamma _{B}+N\right) +f_{T2}e^{-d/2\xi
}\left( \gamma _{B}-N\right) }{e^{-d/\xi }\left( \gamma _{B}-N\right)
^{2}-e^{d/\xi }\left( \gamma _{B}+N\right) ^{2}},  \label{AB}
\end{eqnarray}%
where $N=\operatorname{Re}F\operatorname{Re}F_{N}$.

\begin{figure}[tbp]
\includegraphics [scale=0.70]{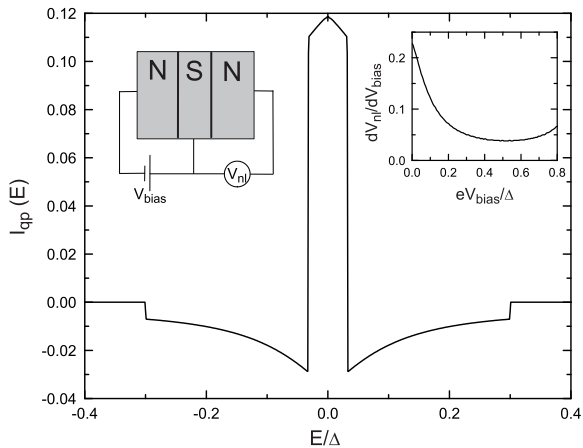}
\caption{Spectral quasiparticle current at zero temperature
accross the second interface under a voltage bias of the first
interface, $V_{\operatorname{bias} }=0.3 \Delta$, and a zero total
current across the second interface as sketched. The
superconducting interlayer thickness $d=0.3\protect\xi$, and $
\protect\gamma_B=5$. Inset: The response of the induced nonlocal
voltage across the second interface (which is the measured
quantity of Ref.~\onlinecite{Klapwijk}) as function of the applied
voltage bias across the first interface.} \label{V1V3}
\end{figure}

The bias condition provides the final equation from which the
distribution functions and currents are derived. The first
interface is biased with a voltage $V_{\operatorname{bias}}$, from
which $f_{T1}$ is then known. Two bias conditions for the second
interface are considered. (a) $I_{qp}=0$: For a zero total current
through the second interface, i.e. an open connection, or an ideal
Voltmeter, the nonlocally induced voltage has to be found
self-consistently from Eq. (\ref{AB}) and the additional boundary
condition,
\begin{equation}
I_{qp}=\int_{-\infty }^{\infty }D_{T}\frac{\partial }{\partial x}f_{T}dE=0.
\notag
\end{equation}
(b) $V_{S}=V_{N_2}$: Under the alternative bias condition of zero
potential difference across the second interface, $f_{T2}=0$ and
Eq. (\ref{AB}) directly provides the solution for $f_{T}$ in the
superconductor.

Note, that it is essential to our approach that
$\operatorname{Re}F_{N}$ is nonzero at subgap energies. When these
proximity induced correlations are neglected, the quasiparticle
current $I_{qp}$ and the nonlocal effects vanish, coinciding with
the results from previous tunnel Hamiltonian \cite{Falci,Bignon}
and circuit theory \cite{Morten} models.

\begin{figure}[tbp]
\includegraphics [scale=0.70]{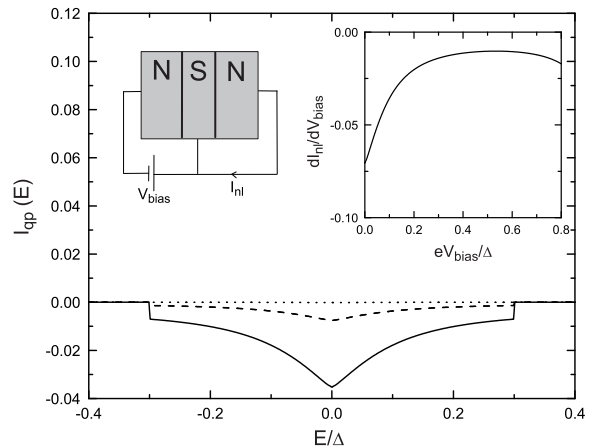}
\caption{Nonlocal spectral quasiparticle current at zero
temperature across the second interface under a voltage bias of
the first interface, $V_{\operatorname{ bias}}=0.3\Delta $, and a
zero voltage bias of the second interface as sketched.
$\protect\gamma _{B}=5$ and $d=0.3\protect\xi $ (solid line), $d=
\protect\xi $ (dashed line), $d=3\protect\xi $ (dotted line).
Inset: The response of the total nonlocal current across the
second interface as function of the applied voltage bias across
the first interface, for $ \protect\gamma _{B}=5$ and
$d=0.3\protect\xi $.} \label{Gnl}
\end{figure}

The resulting spectral quasiparticle currents across the second
interface are shown in Figs.~\ref {V1V3} and \ref{Gnl}, for a thin
superconductor of $d=0.3\xi$ embedded symmetrically between two
tunnel barriers with $\gamma _{B}=5$. In the case of bias
condition (a), a nonlocal voltage, $V_{nl}$, is induced in the
unbiased normal metal electrode, see Fig.~\ref{V1V3}, as was
experimentally observed \cite{Klapwijk}. $V_{nl} $ is the source
of a local Andreev reflection process at the second interface, as
characterized by the low-energy peak in $I_{qp}\left( E\right) $.
In Fig.~\ref{schematic}, also $I_{qp}\left( E\right) $ across the
first interface is shown, to illustrate the different
quasiparticle tunneling and reflection processes. A negative
current corresponds to a flow of electrons in the positive
direction. $I_{qp}\left( E\right) $ at the second interface,
outside the spectral region of Andreev reflection ($\left\vert
E\right\vert>V_{nl}$) is physically caused by the two nonlocal
processes of elastic cotunneling (EC) and crossed Andreev
reflection (CAR). The reverse backflow process of EC and CAR from
the right to the left electrode results in a suppression of
$I_{qp}\left( E\right) $ at the first interface for $ \left\vert
E\right\vert <V_{nl}$.

In bias situation (b), only the nonlocal currents are contained in
$I_{qp}\left( E\right)$, see Fig.~\ref{Gnl}. The magnitude of the
nonlocal currents scales exponentially with $d/\xi$ [see solution
Eq. (\ref{AB}) and Fig.~\ref{Gnl}], as expected from the fact that
a Cooper pair has size $\xi$ and that nonlocal effects exist by
virtue of the coupling to the superconducting condensate. Thus,
the Thouless scale is not relevant to our model. The nonlocal
currents and voltage scale with $\gamma_B^{-2}$, which provides a
much larger effect than was found from the tunnel Hamiltonian
approach (fourth power in transparency) \cite{Falci,Bignon}.

In the linear response regime (low temperature and low voltage),
it was shown recently by Morten \textit{et al.}~\cite{Morten} that
\begin{eqnarray}
\frac {\partial I_{L,qp}}{\partial
V_{\operatorname{bias}}}=G_{\operatorname{EC}}
\left(eV_{\operatorname{bias}}\right)\pm
G_{\operatorname{CAR}}\left(eV_{\operatorname{bias}} \right),
\end{eqnarray}
where $G_{\operatorname{EC}}$ and $G_{\operatorname{CAR}}$ are
defined as the nonlocal conductances for EC and CAR respectively.
From the fact, that $D_L=0$ in the superconductor at subgap
voltage, and Eq. (\ref{IL}), it follows that $I_L=0$, and $
G_{\operatorname{EC}}=-G_{\operatorname{CAR}}$. This means that
the nonlocal quasiparticle current,
$I_{qp}\left(E\right)=-\left(G_{\operatorname{EC}}-G_{\operatorname{CAR}}
\right)f_{T1}$, as shown in Fig.~\ref{Gnl}, is carried by the EC
and CAR processes equally, \textit{i.e.} $
I_{\operatorname{EC}}=I_{\operatorname{CAR}}=I_{qp}\left(E\right)/2$.
The CAR hole in the second electrode can, therefore, be thought of
as moving in the negative direction, while the EC electron moves
in the positive direction (see Fig.~1).

A treatment of the nonlocal processes in terms of electrons and
holes can be derived from the respective electron and hole
distribution functions, $f_e$ and $f_h$, that are given by
\cite{Belzig} $f_{h,e}= \left(1 \pm f_L-f_T \right)/2$, when also
$f_L$ is treated self-consistently.

The sign and magnitudes of the modeled nonlocal effects are of use
for the interpretation of recent experiments
\cite{Klapwijk,Beckmann,Koren}, although many aspects of the
experiment related to the geometry are not covered by our model.
The obtained equal contribution of EC and CAR to nonlocal currents
in $NSN$ structures indicates that additional quasiparticle
manipulations are necessary before the device can be considered as
a useful source of entangled particles. Creating a non-equilibrium
distribution in the electrodes (for example by energy-filtering in
a Fabry-Perot structure) in this respect would be benificial.

\section{Conclusion}
In summary, within the assumptions of quasiclassical
Keldysh-Usadel theory, we find that nonlocal EC and CAR effects
exist in a diffusive quasi-1D wire by virtue of the proximity
effect. We have found that CAR and EC have the \textit{same} sign
in their contribution to the nonlocal spectral quasiparticle
current and nonlocal voltage. CAR and EC scale with the square of
the barrier transparency, providing large nonlocal effects, of the
order of experimentally observed magnitude.

\begin{acknowledgments}
Discussions with V. Chandrasekhar, S. Kawabata, T.M. Klapwijk,
M.Yu. Kupriyanov, A. Morpurgo, F. Pistolesi, and A.D. Zaikin are
gratefully acknowledged. The work was partially supported by the
Netherlands Organisation for Scientific Research (NWO) and the
NanoNed program under grant TCS7029.
\end{acknowledgments}

\end{document}